\documentclass{article}
\usepackage{graphicx}

\begin{document}

\title{Hierarchical Markovian modeling of multi-time systems} 

\author{M. Abel$^{1}$, K.H. Andersen$^{2}$ and G. Lacorata$^{3}$
\vspace{1cm}\\
$^1$ Universit\"at Potsdam, Institut f\"ur Physik, \\Postfach 601553,
D-14415 Potsdam, Germany.\\
$^2$ Dept.~of Mechanical Engineering, Technical University of Denmark,\\
DK-2800 Kgs.~Lyngby.\\
$^3$ Dipartimento di Fisica, Universit\`a dell'Aquila, \\
Via Vetoio 1, I-67010 Coppito, L'Aquila, Italy.
}

\maketitle
\begin{abstract}
\noindent
We present a systematic way to analyze and model systems having many
characteristic time-scales.  The method we propose is employed for a
test-case of a meandering jet model manifesting chaotic tracer
dispersion with long time-correlations.  We first choose a suitable
state space partition and analyze the symbolic dynamics associated to
the fluid particle position.  In a second step we construct a
stochastic process in terms of a multi-time Markovian model. This
corresponds to a hierarchy of random travelers on a graph where each
traveler moves at his own time scale. The results are compared on the
basis of statistical measures such as entropies and correlation
functions.
\end{abstract}

\section{Introduction}
\label{section:intro}
Extended systems, as they are found in nature, often show
complicated dynamics on several characteristic scales and times.   
One, often cited, example is
turbulent motion, where the dynamics stretches from an integral scale
to the dissipation scale.  In many situations, it is possible to find
a low-order approximation of the dynamics and one remains with the
task to describe the temporal behavior of the system under
consideration. A standard example is the area of pattern formation
\cite{Cross-Hohenberg-93} where one ends up with systems of ordinary
differential equations, describing the evolution of a few spatial
modes.  It is, however, not always possible to perform a mode
decomposition, but nevertheless spatial structure is clearly present.  
Then it makes sense to identify states and investigate the
dynamics and interaction of these states. There exists a huge amount
of works concerning deterministic \cite{Kantz-Schreiber-97}, and
stochastic modeling \cite{Gardiner-96}.  Examples for the above
mentioned systems are found among others in climatology
\cite{Storch-Zwiers-99}, oceanography
\cite{Cencini-Lacorata-Vulpiani-Zambianchi-99}, or ecosystems
\cite{Murray-93}. In these examples, one typically follows tracer
particles that are put into the respective flow and performs some
partitioning afterwards. We do not want to bother with this first,
very subtle step of partitioning, rather we take the partition as
given.

On the other hand, there exist systems where one directly observes the
temporal signal of several states, which can be coupled. A partitioning
is then obviously not necessary and one is not bothered anymore with
the problems arising from that point. Examples are systems of coupled
oscillators, coupled ODE's.
Between the two described scenarios lie numerical simulations
where it is often easy to identify spatial states by physical 
\cite{Weisheimer-Dethloff-2001, Corti-Molteni-Palmer-99} or
mathematical  \cite{Coulliette-Wiggins-2001} arguments.

Often, one considers coupled systems with one dominant time scale and
different typical amplitudes. States with fast oscillations are
usually assumed to have small amplitude and are neglected. This way,
small oscillations that might well play a role in inducing finite-size
perturbations and chaos in a system are neglected. On the other hand,
it can well appear that the fast fluctuations have large amplitude and
thus obscure the slow states dynamics. 

Chaoticity of nonlinear  systems implies the loss of memory of a
tracer that moves in the (either extended or discrete) system. It thus
makes sense to use a statistical description. In this article, we
present a way to treat systems with several states, which are
possibly coupled and act on different time scales.  We explain a
method to filter time signals and extract information at different
time scales along with a non-conventional, general method to construct
a stochastic model capable to approximate the evolution of the system
under observation.  We end up with a statistical description of a
multi time scale dynamics in terms of a Markovian cascade process.
The different time scales are identified and the dynamics is
reconstructed in the statistical sense. This is done in order to
compare the reconstructed and original dynamics by means of complexity
measures and correlation functions to ensure that the modeling is
senseful.  Apart from the theoretical interest about how to formulate a
statistical model, we can figure out mainly two practical uses:
firstly, one can implement a statistical model as a building block in
geophysical or other large scale simulations to get an idea of
the fast (and possibly small) scale fluctuations and secondly, on
could use transition times and probabilities between different states
to evaluate the most probable traces a passive scalar (e.g., a
contaminant) would have in the described flow (e.g., the  atmosphere).

In this article we treat an example from geophysics: the chaotic
dispersion process generated from a meandering jet model.  Apart from
the interest in itself, we believe that this system is worth noting
for the following reasons: {\it i)} it shows chaos and mixing on two
time scales as desired by our research goal (and thus represents some
minimal setting giving the possibility for detailed studies of the
methods we developed), {\it ii)} it is typical for geophysical
applications {\it iii)} we can compare results explicitly, since
there exists an older article attempting to solve the task with
conventional methods.

The article is organized as follows: After this brief introduction, we
present in Section \ref{section:model} the model, in Section
\ref{section:entropy-and-filtering} the analysis method is explained,
in Section \ref{section:construction} the construction of the model
trajectory is described, Section \ref{section:results} contains the
results and the comparison of constructed and original signal, finally
we end with a short discussion and some conclusions in  Section
\ref{section:discussion}.

\section{The model: a signal with different characteristic times} 
\label{section:model}

\begin{figure}[tb]
  \begin{center}
  \includegraphics[width=0.45\textwidth]{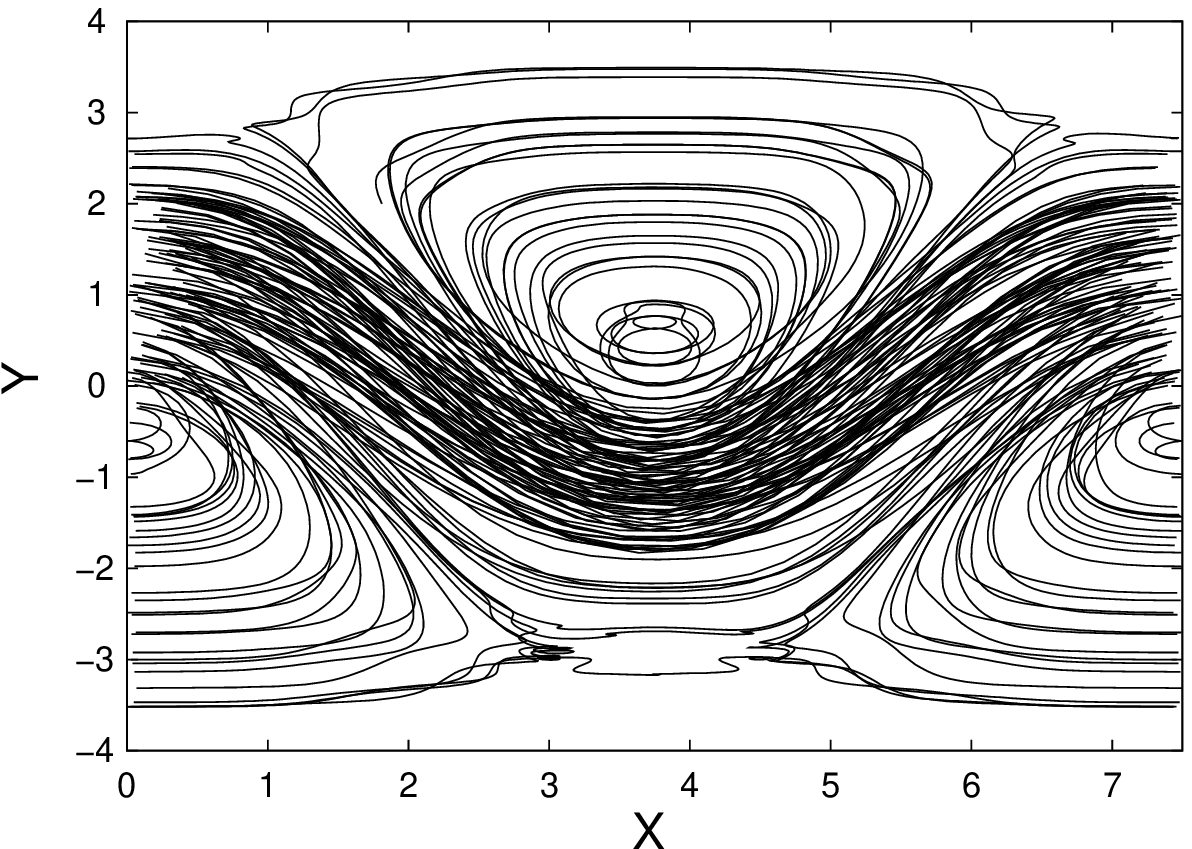}
  \includegraphics[width=0.45\textwidth]{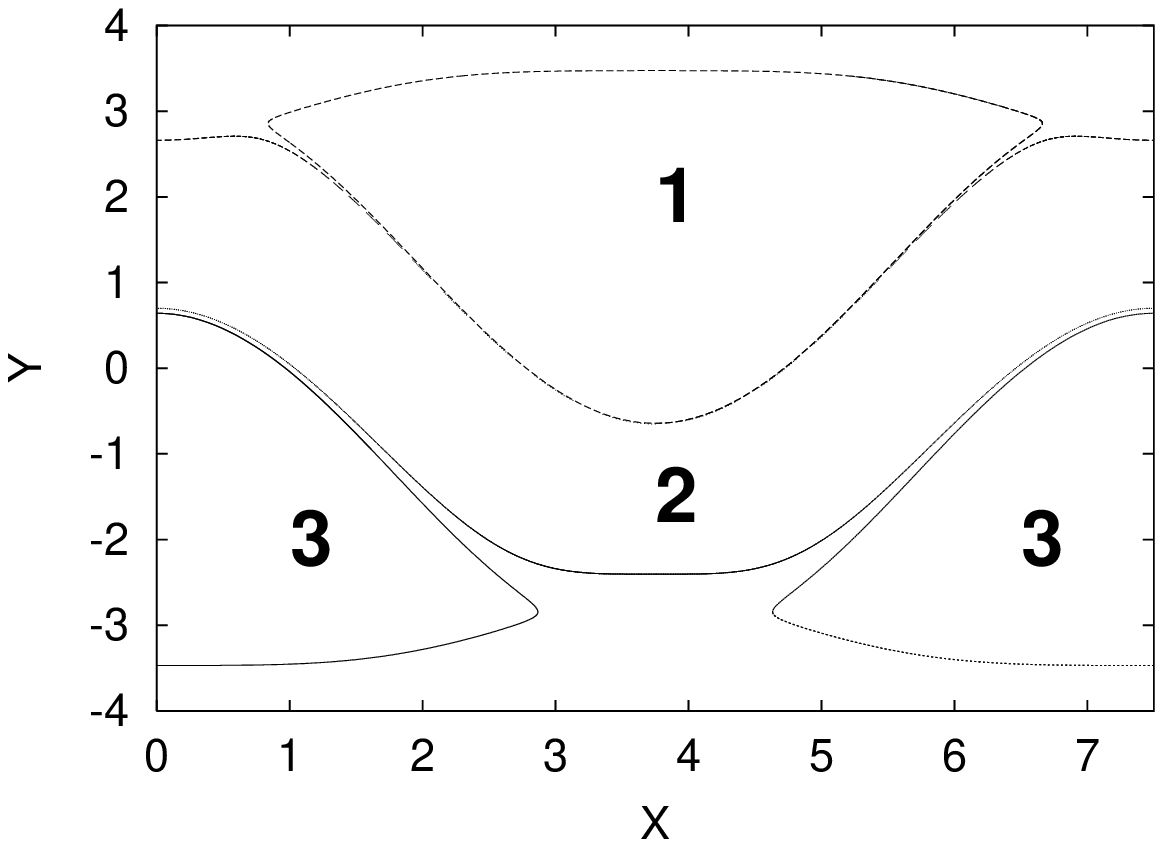}
  \caption{Left: snapshot of the streamlines in the meandering jet model
    (\protect\ref{eq:psi}). The gyres are seen as closed orbits to the
    north and south of the jet. The boundary between the jet and the
    gyres, as defined here, correspond to $\Psi = \pm 0.60$. Spatial
    coordinates are measured in units of the wavelength $2\pi/k$.
    Right: a particle trajectory (periodic boundary conditions). In
    the corresponding gulf stream system the particle experiences an
    average drift from the left to the right.  The tracer can be
    caught by a gyre (states 1 and 3) or the jet (state 2) for a
    longer time (dense tracks).}
  \label{fig:stream}
  \end{center}
\end{figure}

Let us consider a simple but non-trivial case of multi-time dynamics
in a geophysical system: Lagrangian transport across a meandering jet,
formerly introduced as kinematic model of the Gulf stream by Bower
\cite{Bower-91} and Samelson \cite{Samelson-92}.  The model serves our
purposes, since it can be well used as the test case in which only two
characteristic times are involved as we will see below (i.e.  fast and
slow particle transitions across the flow).  In this model, the gulf
stream is represented by a central meandering jet, flanked by gyres
rotating in opposite directions (Fig.~\ref{fig:stream}).  The model is
described by the stream function \cite{Bower-91,Samelson-92}:
\begin{equation}
  \Psi(x,y) = -\tanh {y - B_0 \cos(kx) \over \sqrt{1 + B_0^2 k^2 \sin^2(kx)}} 
  + cy,
  \label{eq:psi}
\end{equation}
where $x$ and $y$ are the spatial coordinates of a fluid particle,  
$k$ is the spatial wave number of the meander structure, $B_0$
controls the amplitude of the meanders, the term in the denominator of
(\ref{eq:psi}) defines the width of the jet and $c$ is the velocity in
the ``far field'' north and south of the (westerly) jet current. The
fluid particle velocity components $(u,v)$ at the point $(x,y)$ are:
\begin{equation}
  u = - {\partial \Psi \over \partial y}, \quad \mathrm{and}\quad
  v = {\partial \Psi \over \partial x}.
  \label{eq:vel}
\end{equation}

Chaos is introduced by adding an explicit time
dependence to the stream function (\ref{eq:psi}) by replacing
$B_0$ with $B(t)$ defined as
\begin{equation}
  B(t)=B_0 + \epsilon \cdot \cos(\omega t + \theta)
  \label{eq:B}
\end{equation}
This induces a periodic oscillation of the meander, with period $T=2
\pi /\omega$, amplitude $\epsilon$ and initial phase $\theta$, around
the mean value $B_0$. Depending on the values of the perturbation
parameters, $\omega$ and $\epsilon$, chaotic motion occurs in the
vicinity of the separatrices. For appropriate values of the perturbation
 parameters, resonance overlap  \cite{Chirikov-79} occurs, and
transport between northern and southern gyre can take place.
  
A study of Lagrangian trajectories in one elementary cell of the
streamline pattern shows that there are two basic time scales:
$\tau_g$ which is the advection time of a particle traveling along the
jet core, and $\tau_o = 2\pi/\omega$ which is the period of the
oscillation of the meander.  In the following, in order to obtain
non-dimensional equations we rescale time with $\tau_o$.  The particle position
is sampled accordingly  at the end of each period.

A coarse graining of the space can be obtained by dividing the flow
domain into {\em elements}, e.g., separating regions with closed from
regions with open streamlines.  The partition to be used throughout
this article is defined as: 1) the northern gyre, 2) the jet, 3) the
southern gyre (cf. Fig.~\ref{fig:stream}).  Assigning each of the
partition a number yields a discrete description of the trajectory.

An example of a particle trajectory is seen in
Fig.~\ref{fig:stream}.  For this example (and the rest of this
article) the parameters of the models have been chosen as: meander
wavelength $l_0 = 2 \pi / k = 7.5$, the (mean) meander amplitude
$B_0=1.2$ and $c=0.12$. An example of the corresponding trajectory in
terms of the partitions is seen in Figure \ref{fig:signal_KW}, top
line. The trajectory consists of fast oscillations between one of the
gyres and the jet, interleaved by periods where it stays in one of the
gyres for a longer time. The fast oscillations are basically due to
the fact that the particle undergoes a large number of crossings back
and forth across the boundary of two partitions.
 
\begin{figure}[tbp]
  \begin{center}
    \includegraphics[width=0.75\textwidth]{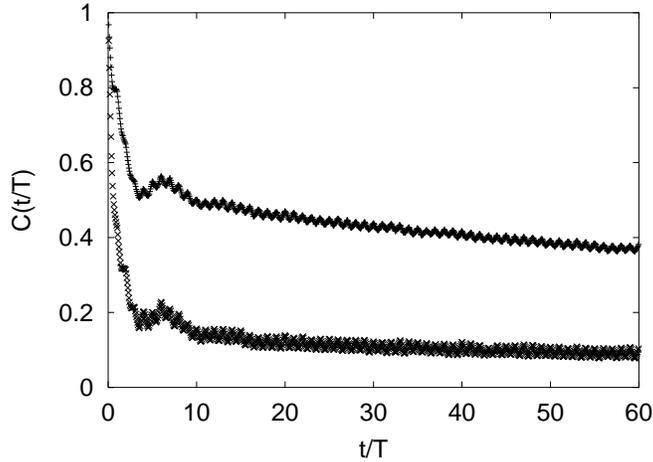}
    \caption{Correlation function of the signal shown in
    Fig.~\ref{fig:signal_KW}, top graph. The long-time correlations are clearly
    present but cannot be captured by a usual Markov chain with
    equidistant time steps.}
\label{fig:cf-bare} \end{center}
\end{figure}

\section{Entropic analysis and filtering}
\label{section:entropy-and-filtering}
The aim of this section is the presentation of a suitable filtering
technique which allows to highlight and characterize the dynamical
features at different characteristic times. This filtering method,
developed on the basis of an exit time approach to entropy
computation, is a way to understand how to find a good stochastic
model for the process.

\subsection{Encoding of the signal}
Let us consider the time signal of an observable $s(t)$.  Typically
$s(t)$ is registered at a certain sampling rate $\tau_s^{-1}$, i.e.,
we consider a discrete time evolution, $s(n)$ with $t=n \cdot \tau_s$.
A coarse-grained description of $s(n)$ is obtained by dividing the
range of $s(n)$ through a partition with, say, $P$ elements, and
assigning a label to each element of the partition, say $\sigma_1,
\sigma_2, ..., \sigma_P$.  Consequently, the time evolution of $s(n)$
is mapped into a symbolic dynamics $\sigma(n)$, with $\sigma(n) \in
\{\sigma_1, \sigma_2, ..., \sigma_P\}$.  Let us indicate with $\sigma$
a generic symbol of the sequence and with $\tau$ the lifetime of
$\sigma$, i.e., the time the system takes to move from the state
$\sigma$ into a new one. We will use sometimes the terms residence
time or exit time as a synonym for ``lifetime'' when more suitable, in
the context of markov chains, the term waiting time is used (mainly in
the literature of queuing systems).

A sequence of length $n$ is designated as follows:
\begin{eqnarray}
\label{eq:encoding}
  S_n &=& (x_1, x_2, \dots , x_n)\, , \quad \mathrm{where} \\
  x_i &=& (\sigma_i, \tau_i)\, .
\end{eqnarray}
For example, imagine a signal $s(t)=(1,1,1,2,2,2,2,3,3,1,1,...)$ which
assumes positive integer values at each unit time step ($\tau_s=1$);
then the corresponding $(\sigma,\tau)$ variables are
$\sigma_i=(1,2,3,1,...)$ and $\tau_i=(3,4,2,2,...)$, respectively.

\subsection{Estimating Entropies}

Let us briefly remind some recent results concerning the
development of the estimation of entropy using the exit-time approach 
\cite{ABCFVV-2000a,ABCFVV-2000b}. To illustrate the basic
ideas we  describe the method and afterwards discuss its
application to the problem under consideration, some details are
explained in the appendix.

Out of the set of sequences ${S_n}$ we collect statistics for each
subsequence $S_n$ and construct the probability distribution $P(S_n)$.
Now, one can compute the Shannon entropy $h$ of the signal $\sigma$ by
the following identity:
\begin{equation}
h = {h^* \over \langle \tau \rangle}
\label{eq:h}
\end{equation}
where $h^*$ is the Shannon entropy of $x_n$, $h^*=\lim_{n \to \infty}
H^*_{n+1}-H^*_{n}$ (the information production with $n$), with
\begin{equation}
 H^*_n = -\sum_{S_n} P(S_n)\ln P(S_n),
\end{equation}
and $\langle \tau \rangle$, the average life time, is computed as 
\begin{equation}
\langle \tau \rangle = \lim_{N \to \infty} {1 \over N} \sum_{i=1}^{N} \tau_i
\label{eq:meantau}
\end{equation}
Denoting the entropies of the $\sigma$ and $\tau$ sequences with
$h^*(\sigma)$ and $h^*(\tau)$, respectively, one can determine upper
and lower bounds on the entropy:
\begin{equation}
 \max \left[ h(\sigma), h(\tau)\right] < h(\sigma, \tau) < h(\sigma) + h(\tau)
 \label{bounds}
\end{equation}
The details of the method are described in 
%%Appendix A and 
\cite{ABCFVV-2000a,ABCFVV-2000b}. 

\subsection{The filtering procedure}
\begin{figure}[htbp]
  \begin{center}
    \includegraphics[width=0.75\textwidth]{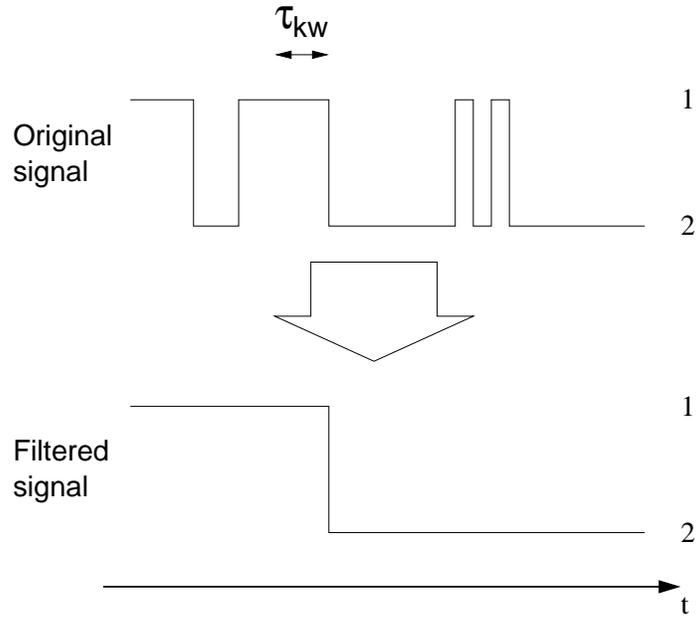}
    \caption{An illustration of how the killing window algorithm
      changes the original signal (top) to a filtered signal
      (bottom). In this example the signal is only alternating between
      two states, 1 and 2.} 
    \label{fig:kw}
  \end{center}
\end{figure}

\begin{figure}[tbp]
  \begin{center}
  \includegraphics[width=0.75\textwidth]{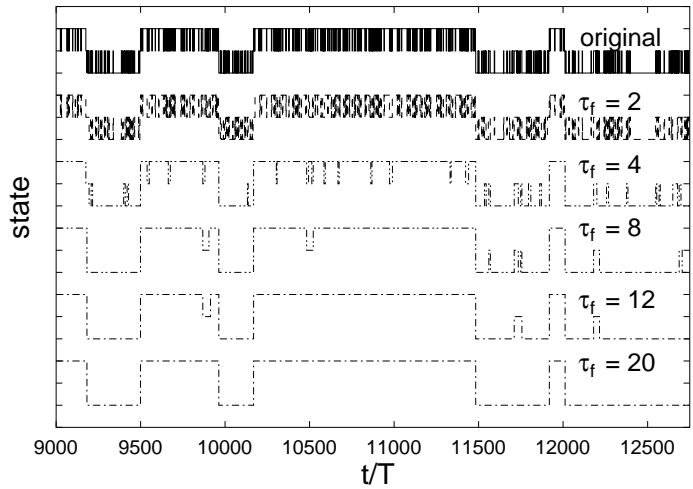}
  \caption{Comparison of signals filtered with different killing
  windows, $\tau_F=0,2,4,8,12,20$, from top to bottom. It is
  evident how the fast fluctuations are removed by the killing
  window. For a very long killing window, $\tau_F=20$, also the
  residence in the jet is killed, and the model has effectively only
  two states.}  
  \label{fig:signal_KW} 
 \end{center}
\end{figure}

Let us now introduce a procedure to select the temporal behavior
``slower than a given frequency''. The filter should therefore be able
to discard the many fast fluctuations from one state to another which
occur very frequently (see figure \ref{fig:signal_KW}, top line).  We
call this filter ``killing window''; it operates in the following
way. If the signal  fluctuates from one state into another
state before returning to the first state, and if this fluctuation
lasts for a shorter time than $\tau_F$ (the filter length of the killing
window), this fluctuation is ignored. Written symbolically this means
that:
\begin{equation}
  (\sigma_1, \tau_1), (\sigma_2, \tau_2), (\sigma_1, \tau_3) \rightarrow 
(\sigma_1, \tau_1+\tau_2+\tau_3)\, ,
\end{equation}
provided that $\tau_2 < \tau_F$. The effect of
the algorithm is shown in Fig.~\ref{fig:kw}, and the role of different
lengths of killing windows is shown in Fig.~\ref{fig:signal_KW}. This
filter has the effect of ``killing'' all fluctuations smaller than
$\tau_F$. Note that is also effectively prolongs the slower 
oscillations, c.f.~Fig.~\ref{fig:kw}. 

With this filtering procedure applied to the original signal $x_n$, we
obtain a filtered sequence $x_{n}^{\tau_F}$, for which we can compute
the entropy $h(\tau_F)$ using (\ref{eq:h}).  We used for simplicity a
filter that acts backwards in time. It is however not difficult to
redefine the filter such as to be invariant under time reversal, the
result remains the same within an error of the fast time scale. The
observed signal varies over two very distant characteristic time
scales, $\tau_{fast}$ and $\tau_{slow}$, respectively. Thus,
information at the slower time scale can be extracted if $\tau_{slow}
\gg \tau_F\; ^>\!\!\!\!_\sim \,\tau_{fast}$, because due to the second
inequality almost all fast oscillations are killed.

\begin{figure}[tbp]
  \begin{center}
    \includegraphics[width=0.5\textwidth]{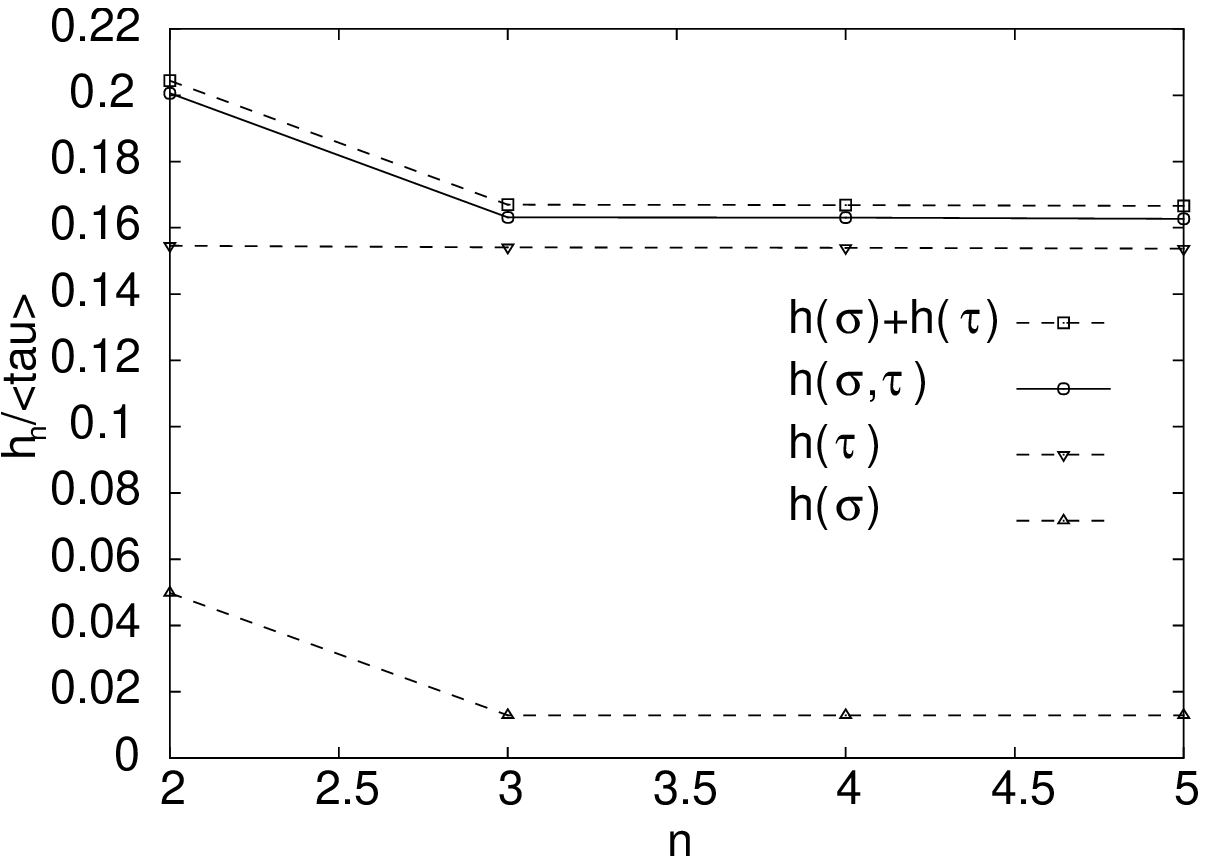}%
    \includegraphics[width=0.5\textwidth]{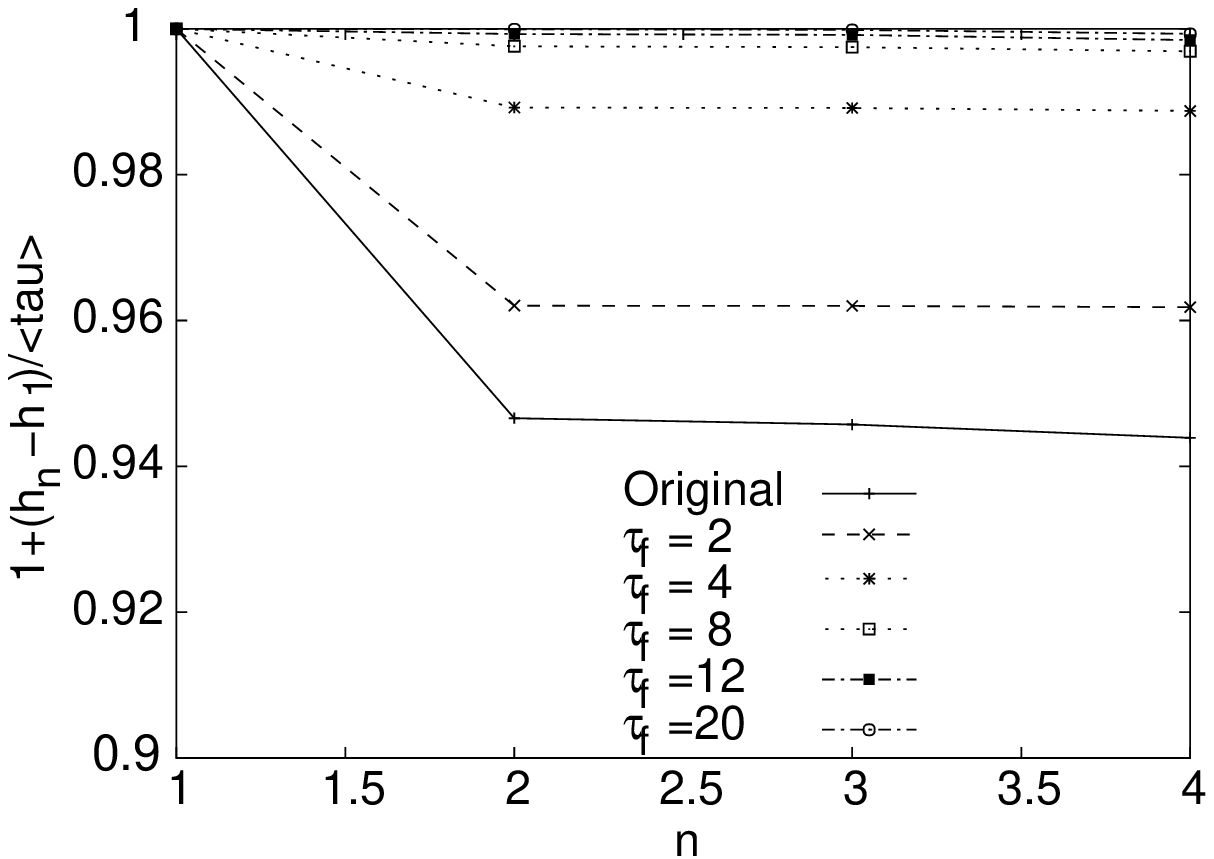}
    \caption{Calculation of the entropies. a) the entropy for
    $\tau_F=2$ together with the upper and lower bounds form
    (\protect\ref{bounds}). b) Comparison of the entropy for varying
    widths of the killing window. The entropies are normalized by
    adding a constant such that $h_1 = 1$.}
    \label{fig:h_n-comp}
  \end{center}
\end{figure}

\begin{figure}[tbp]
  \begin{center}
    \includegraphics[width=0.5\textwidth]{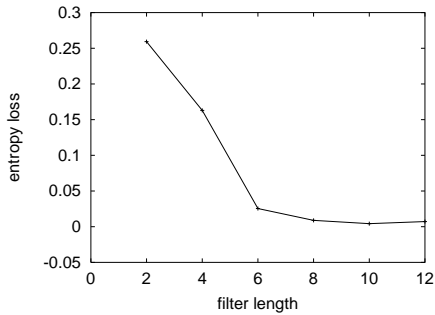}
    \caption{Information loss with filter length. From $\tau_F=8$ we
    have a constant rate of information loss and thus the difference
    between the different filter lengths is zero. This in turn
    justifies the description of the process by one scale only.} 
\label{fig:h-vs-kw} \end{center}
\end{figure}

In Fig.~\ref{fig:h_n-comp}a the convergence of the entropy, at varying
the block size, for a killing window of length 2 is shown, together
with the upper and lower bounds from (\ref{bounds}). First of all we
see that the entropy lies nicely in between the upper and the lower
bounds.  For a killing window of length 2 this is not surprising, as
most of the transitions are fast oscillations between jet and
gyre. However, this entropy converges after $n=2$, which means that one
needs a second order Markovian process in order to properly describe
this process. Let us stress that now the state of the system is
given by $\sigma$ and $\tau$, therefore a Markov model of order $m$
corresponds, in the original space of the symbols ${\sigma}$, to a
Markov model of order $m \cdot \langle \tau/\tau_s \rangle$.  The entropy
based only on the residence times, $h(\tau)$, is larger, and thus
constitutes the effective lower bound for the total entropy.  In
Fig.~\ref{fig:h_n-comp}b this is shown for the bare signal, and for
killing windows of a length from 2 to 20. This entropy based on the
residence times has converged immediately, which means that the
residence times are basically uncorrelated. This is a sign for an
exponential distribution to be confirmed below. 
Now, we have to decide the filter length for which the signal is
described by a process on one time scale only. The killing window
clearly destroys information. If the loss of information is
proportional to the filter length, obviously the process is
homogeneous (with respect to information loss) within the time scale
we are considering.  This is analogous to the information production
rate which has to converge to a constant to find the entropy of a
chaotic system.  \cite{note}. In Fig.~\ref{fig:h-vs-kw} we display the
entropy loss depending on the filter length. For $\tau_f=8$ we have
certainly reached convergence and thus the process can be described at
the level of one single scale (namely the slow one). For completeness,
we remark that almost all information is destroyed if the filter
length exceeds the typical slow time scale, $\tau_F\; ^>\!\!\!\!_\sim
\,\tau_{slow}$.

\section{Construction of the multiple time Markovian Model}
\label{section:construction}
In this section we will demonstrate how the analysis of the signal can
be applied to construct a Markovian model, which can generate
stochastic symbolic sequences with the same statistics as the original
signal. The model should be able to reproduce both the slow and fast
transitions. This is accomplished by first reproducing the slow scale
signal, and then nesting the fast transitions into that one.  We thus
define a ``horizontal'' and a ``vertical'' dependence, the first
describes the process on a specific time scale (slow or fast), the
latter models the interdependencies between the different time scales.
Both are assumed to be Markovian of first order, as our preceding
analysis has shown, a generalization is straightforward. In the
following, for the sake of clarity, we will use capital letters for
slow scale and small ones for the fast scale.

\subsection{Slow transitions}
As determined by the Entropic analysis, the slow signal is the one
filtered by a killing window of length $\geq 8$. The construction
rules are the following: 
\\i) Given an entering state $\sigma_I$ the process
selects according to the transition probability matrix $W_{IJ}$ its
next state $\sigma_J$.
\\ii)Given that selection of the transition $IJ$, the time spent in
$I$ before jumping to $J$ is given by the random variable $\tau_I$,
which in turn possesses the distribution $P(\tau_I)$.

Both needed informations are obtained from the filtered signal.  By
the given rules, we recognize that the construction concentrates on
the {\it transitions} between states, rather than on the state
itself. The {\it lifetime} of the states depends only on this
transition, we see that the transition matrix does not include
self-loops, like $I\rightarrow I$ and thus has zeroes on the
diagonal. In the case of independent lifetimes, i.e., exponential
distribution, one can interchange steps i) and ii) and first select a
state. Numerical examples for the matrices and residence times are
given  Section \ref{section:results}.

 The reconstruction corresponds to a random traveler on a
graph, which stays in each node with a certain node-characteristic
time and then jumps over an edge to the next node according to a
prescribed hopping probability.  
%This picture is visualized in Fig.~\ref{fig:graph}.

\subsection{Nesting of the fast scale}
Up to now, the construction is rather standard. A new thing we
introduce is the vertical connection between scales.  The dynamics of
the fast scale shall depend on the slow one, but apart from that be a
first order Markovian process as described above by the rules i) and
ii). The two quantities we have to consider are the pdf of the
lifetimes $p(\tau_i)$ and the transition probabilities $w_{ij}$. Both
can be quite different from the slow processes quantities.

One obvious condition on the lifetimes is, that the duration of
the fast states time intervals should not exceed the slow ones.  This
is fulfilled, since we anticipated a clear separation of time
scales. In general, the transitions will be hit with an uncertainty of
$ \left<\tau_{fast}\right> \simeq \tau_F $. This in accordance with the
filtering analysis, which is only accurate up to the filter length.
The lifetime distribution function $p$ can be taken without any
restrictions from the analysis, since we anticipate the times to be
independent from each other (as confirmed for our example, see next
section).  The transition probabilities $w_{ij}$, however, have to be
conditioned on the slow scale transitions and thus we have to write
them for a complete description as $w_{ij}(IJ)$.  Results are given in
the next section.

We want to remark, again, that a first order in the space ${\sigma,
\tau}$ corresponds to a very high order in the space ${\sigma}$ with
conventional time sampling. For our illustration by the traveler on a
graph we imagine the traveler moving with his own, slow speed, being
accompanied by his dog which moves forth and back between the present
node of the traveler and some other nodes he's allowed to visit.  The
dog rests only for short times in the nodes, and thus effectively
moves at a fast time scale.  

In conclusion, we have to find for each of the transitions $IJ$ the
according fast transition probability $w_{ij}(IJ)$. The fast process
evolves then with the rules from $w_{ij}(IJ)$ and the lifetime
statistics $p(\tau)$ within the window between two transitions.

In a generalized model, $N$ different time scales are involved, then we
order the processes from slowest to fastest time and construct the
hierarchy by nesting the scale $n$ into scale $n-1$. This
generalization is straightforward and will be reported elsewhere.
Another generalization concerns Markov processes of higher order. This
requires  accurate bookkeeping of the dependencies but in
principle is straightforward, too.

\section{Results}
\label{section:results}
In this section we will present how the technique described above
compares to the real signal. The slow scale transition matrix, we
obtain for our model
\begin{equation}
\label{eq:slow-matrix}
  W_{IJ} = \left( \begin{array}{ccc}
      0 & 0.65 & 0.35 \\
      0.5 & 0 & 0.5 \\
      0.35 & 0.65 & 0 \\ \end{array} \right).
\end{equation}
Note that due the symmetries of the meandering jet problem, this
matrix could in principle be reduced to only its upper half. 
In general, this is
however not  the case.  Since we have a very simple system
with strong restrictions the fast transition matrices look simple,
e.g., for $w_{ij}(12)$:
\begin{equation}
\label{eq:fast-matrix}
  w_{ij}(12) = \left( \begin{array}{ccc}
      0 & 1 & 0 \\
      1 & 0 & 0 \\
      0 & 0 & 0 \\ \end{array} \right).
\end{equation}

\begin{figure}[tbp]
  \begin{center}
  \includegraphics[width=0.4\textwidth]{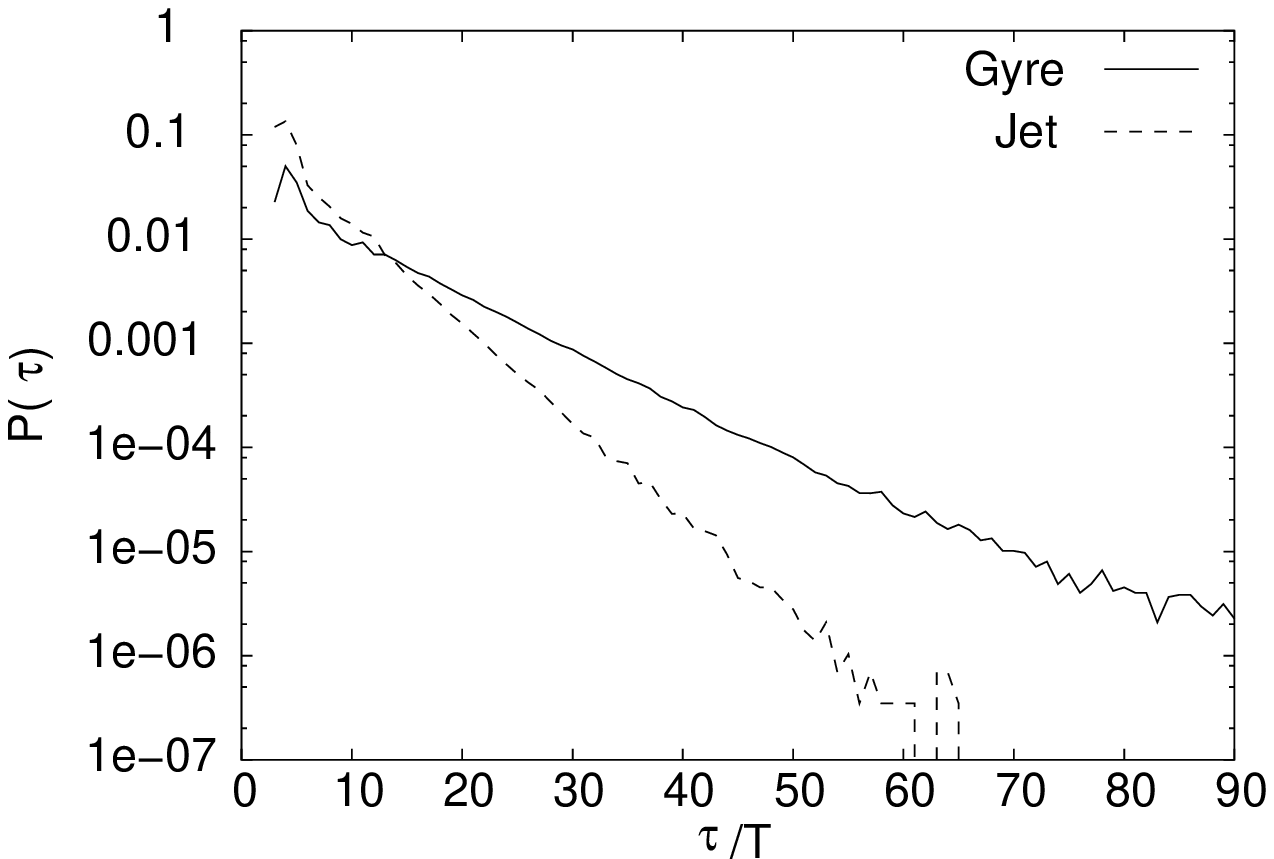}%
  \includegraphics[width=0.4\textwidth]{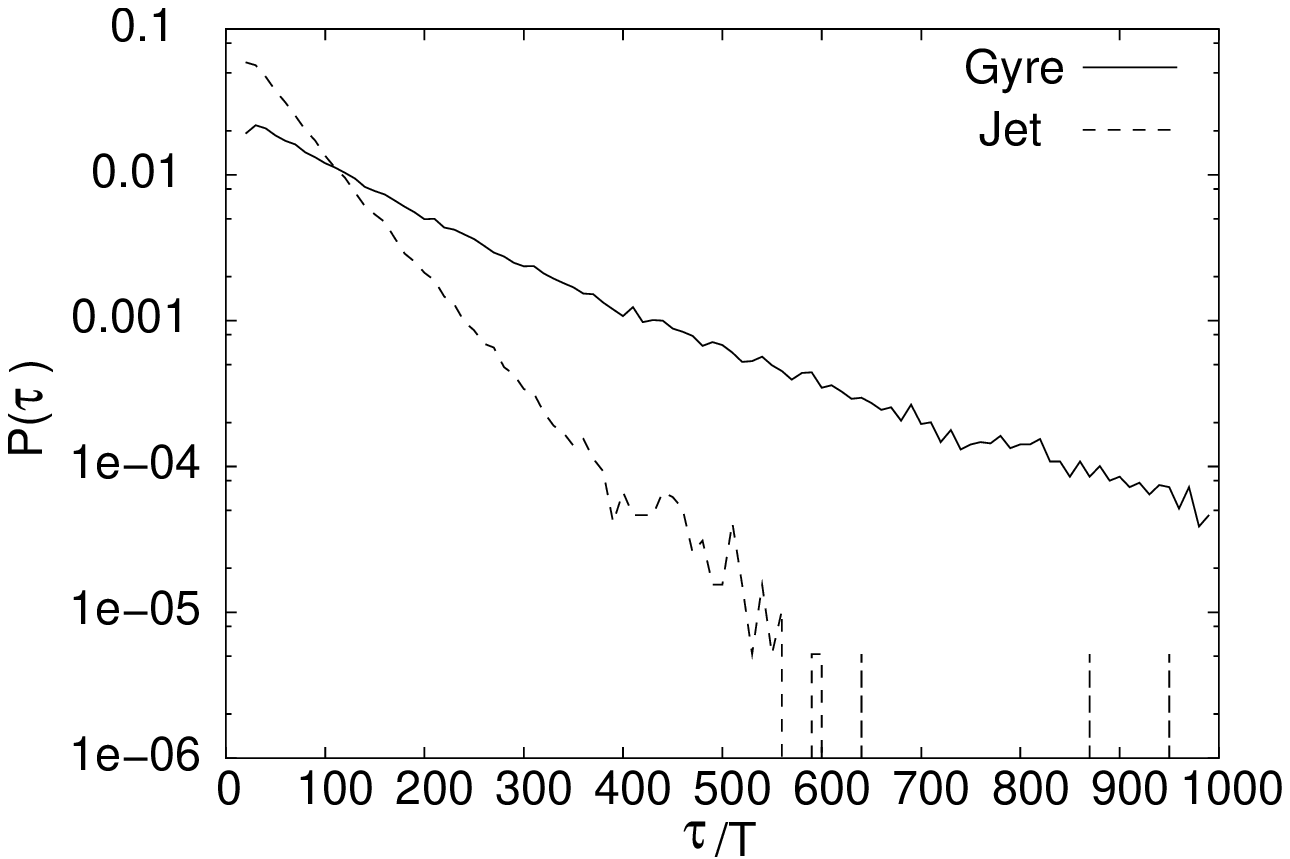}
  \caption{The pdf of the residence times in the jet and the gyres,
  calculated with a killing window of length 2 (left) and 12 (right).}
  \label{fig:pdfs} 
 \end{center}   
\end{figure}

The residence times for the slow and fast scale are displayed in
Fig.~\ref{fig:pdfs}.
Clearly they are very close to be exponential.  This is a sign of a
non-correlated process, which has also been found by the entropic analysis.
From a fit to an exponential distribution, we determine the mean
residence time for the slow time scale to be $\left< \tau
\right>_{slow,gyre}= 64.5$ and $\left< \tau \right>_{slow,jet}=90.9$,
for the fast time scale we obtain $\left< \tau
\right>_{fast,gyre}=5.21$ and $\left< \tau \right>_{fast,jet}=5.83$,
respectively.

%To clarify our procedure, we will spend a paragraph on the
%interpretation of the whole procedure.  
For our example of a coarse grained spatial system, the trajectory,
basically slowly drifts forth and back between north and south gyre,
crossing the jet. During this evolution, fluctuations occur about the
boundaries of the states.  Take, e.g.,~a trajectory starting at the
northern gyre. Having made the transition to the jet on the slow
scale, still, fast fluctuations will occur between the gyre and the
jet, thus masking the signal as in Fig.~\ref{fig:signal_KW}.  If the
trajectory evolves towards the southern gyre, the fluctuations will
happen only between southern gyre and jet.

This does not at all conflict with a probabilistic interpretation: The
signal chooses randomly, according to the transition matrix and
lifetime distribution a new state and lifetime. Information about
history and details of the space evolution are forgotten (up to 1st
order). This is the natural result of the coarse graining. The fast
fluctuations are again random and follow their description with
matrices and lifetimes. In our case, we find the trivial transition
matrices, like in Eq.~(\ref{eq:fast-matrix}) due to the simplicity of
the meandering jet system. In general, however,  more
complicated scenarios will appear.

\begin{figure}[tbp]
 \begin{center}
  \includegraphics[width=0.8\textwidth,angle=0]{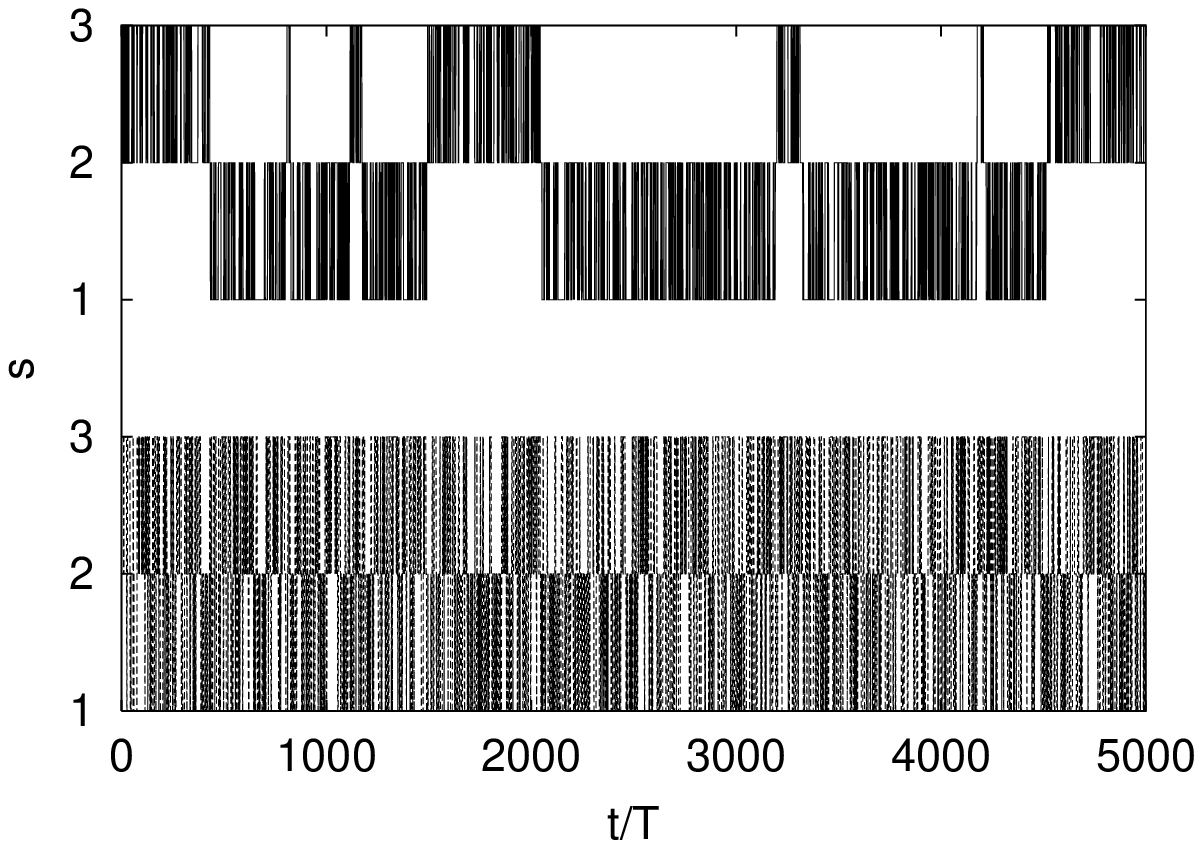}
  \includegraphics[width=0.8\textwidth,angle=0]{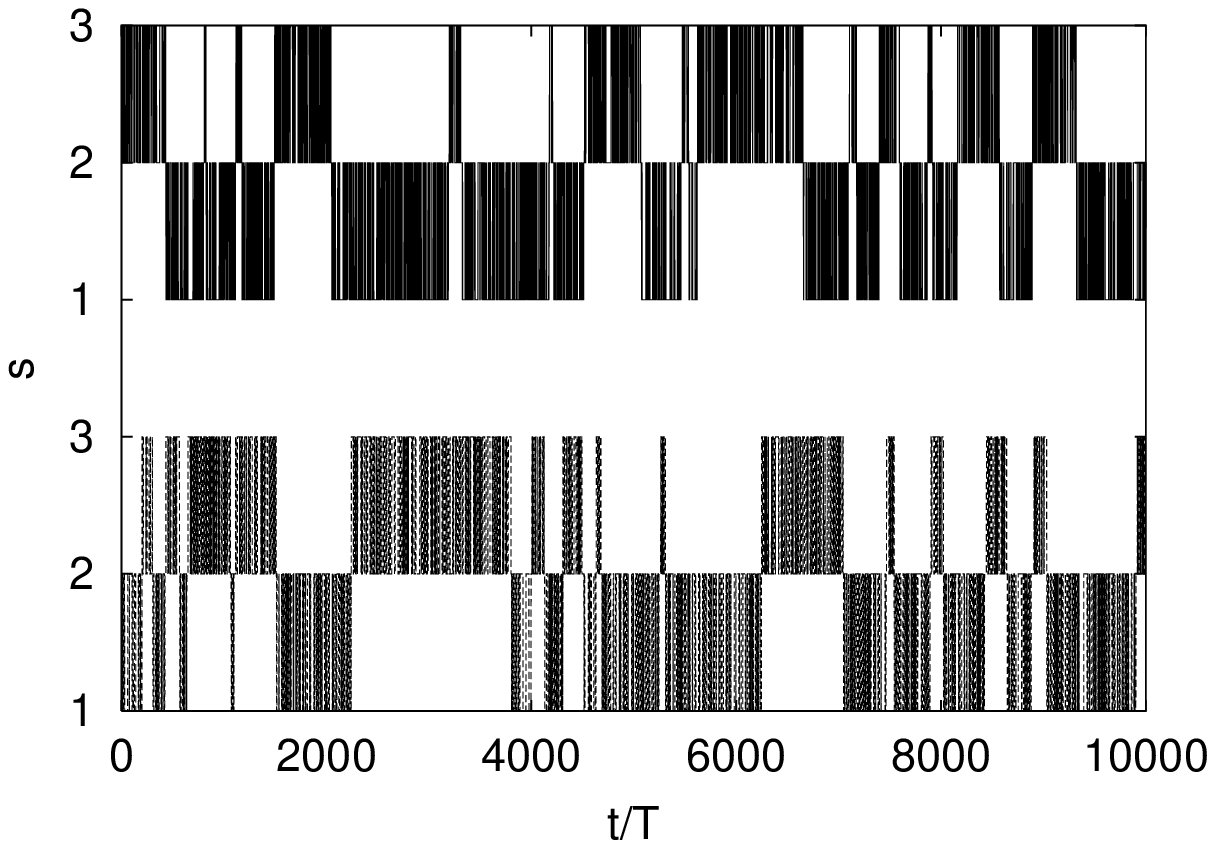}
  \caption{Comparison of the reconstructed signal with the true
  one. a) conventional construction, using fixed sampling time
  (cf. \cite{Cencini-Lacorata-Vulpiani-Zambianchi-99}). Upper part:
  original, lower part: construction. The slow scale transitions are
  not recovered and the signals differ a lot.  b) signal as
  reconstructed by the Markovian cascade. Upper part: original, lower
  part: reconstruction.  It is not obvious at once, which signal is
  the original one.}  
\label{fig:signal-comp} 
\end{center}
\end{figure}

%With the constructed signals, the first comparison is done bye eye
%which has proved to be a helpful tool in many situations. 
In Fig.~\ref{fig:signal-comp}b we show the signal for $\tau_f= 2$ and a
model trajectory for the same parameters, as constructed by our
method.  In contrast to an older work
\cite{Cencini-Lacorata-Vulpiani-Zambianchi-99} the signals resemble
each other a lot, for comparison these results are reproduced in 
Fig.~\ref{fig:signal-comp}a.

\begin{figure}[tbp]
  \begin{center}
  \includegraphics[width=0.75\textwidth]{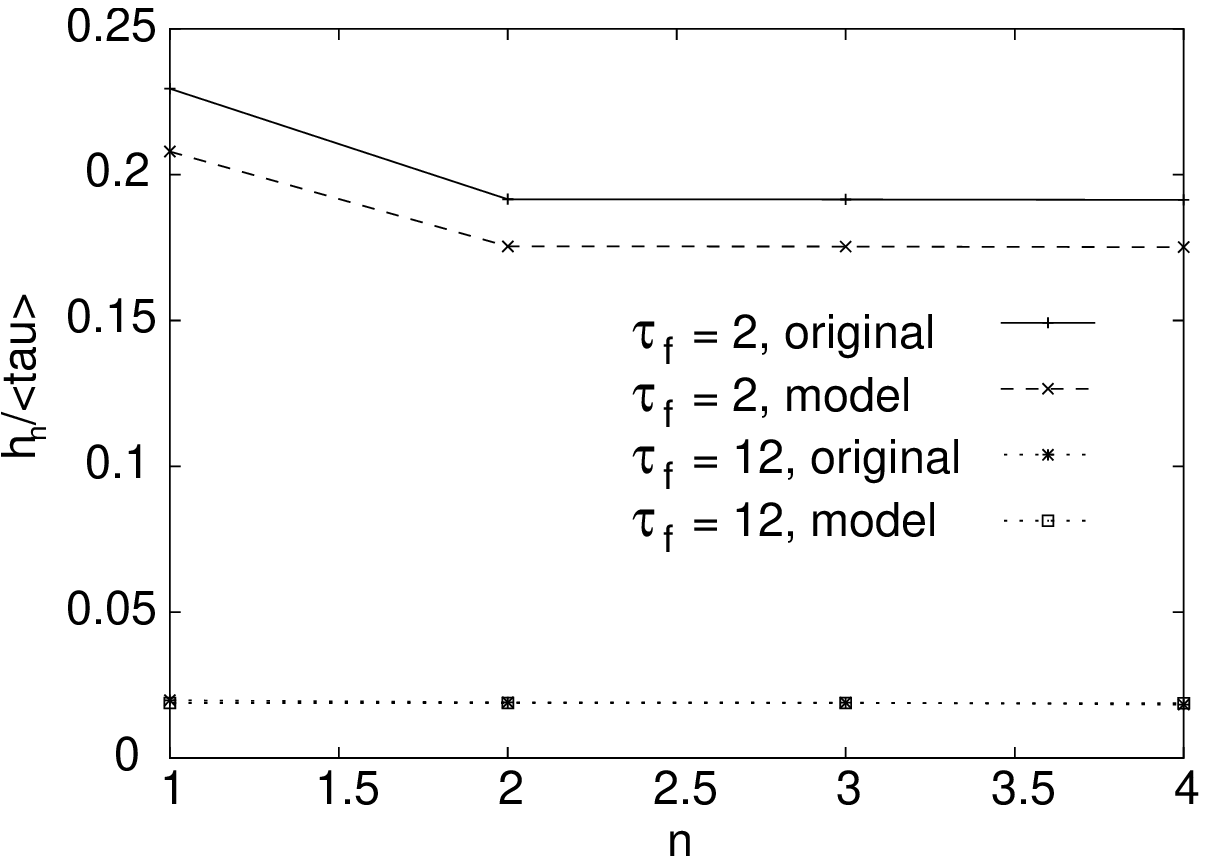}
  \caption{Comparison of the entropies calculated for the original signal
  and for the stochastic model.}
\label{fig:entr-comp} 
\end{center}
\end{figure}

Now we present some more quantitative comparisons. The first quantity
to check is the information contained in the signals, it should be
approximately the same, especially if we consider that this has been
the main tool to find the models order. In Fig.~\ref{fig:entr-comp}
the comparison for $\tau_f=2$ and $\tau_f=12$ is shown -- we find an excellent
agreement.

\begin{figure}[tbp]
  \begin{center} 
     \includegraphics[width=0.5\textwidth]{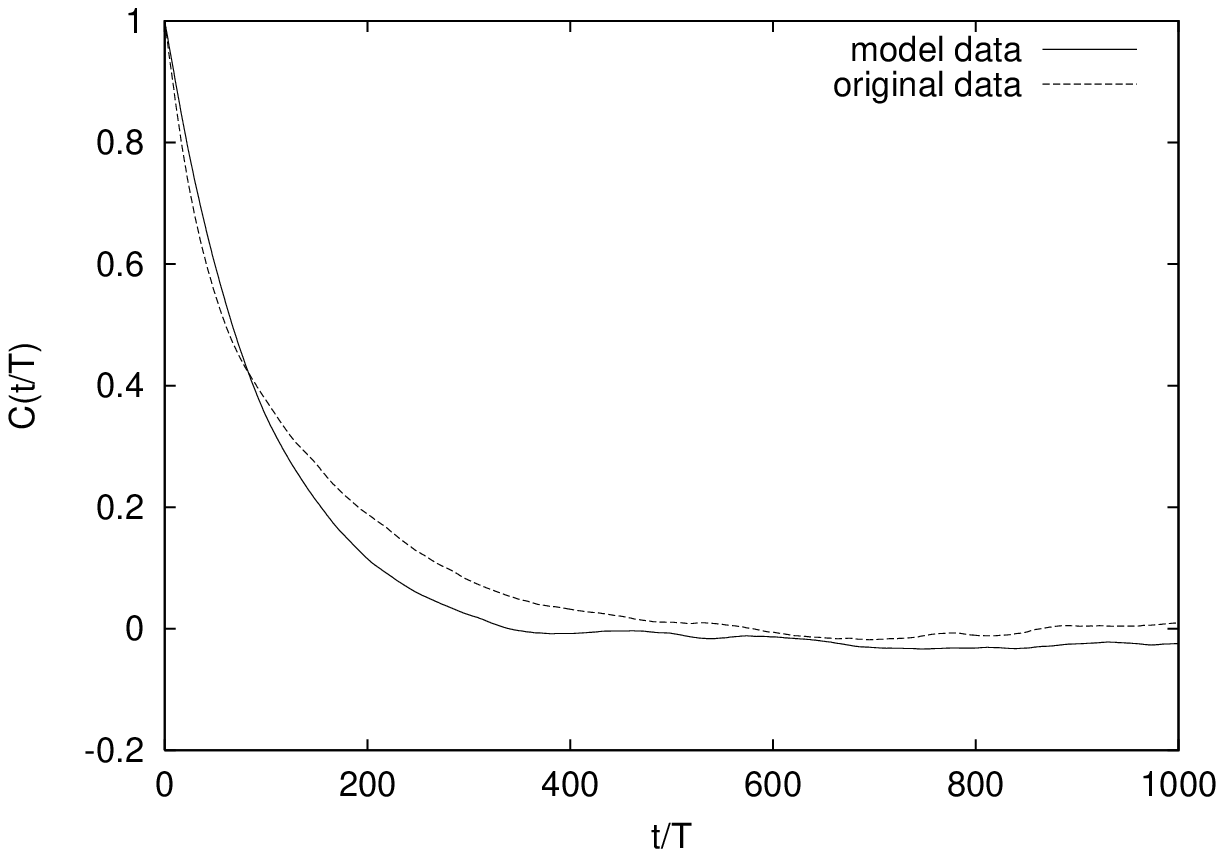}%
     \includegraphics[width=0.5\textwidth]{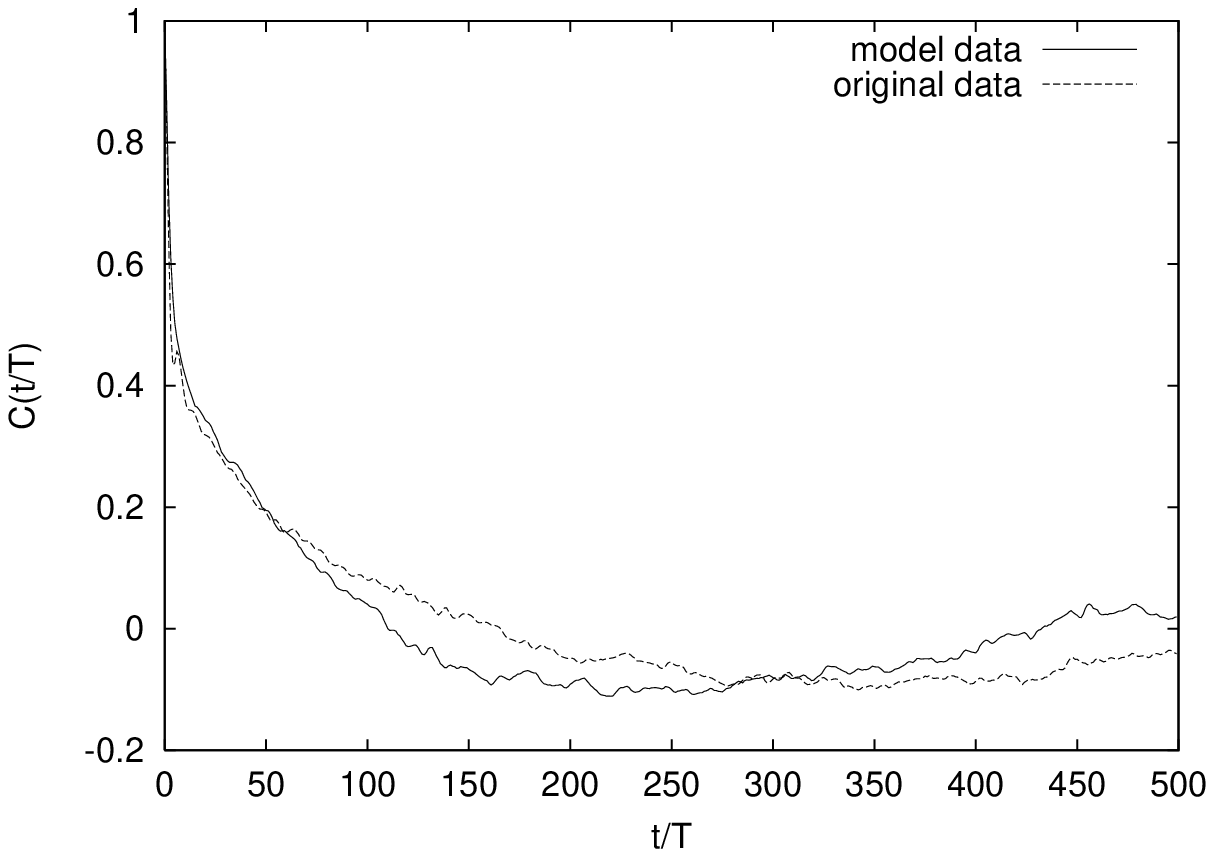} 
  \caption{Left: correlation function for the filtered signal with
  $\tau_F=12$ and the corresponding reconstructed signal. Right: the
  same for $\tau_F=2$.  The rapid decay of the correlations is
  estimated very well by the model data.}
\label{fig:CF-comp} 
\end{center}
\end{figure}

Finally we check for the correlation function as the standard tool to
investigate dependencies within the signal. The result is displayed in
Fig.~\ref{fig:CF-comp}. The general behavior for the correlation
functions is the same. It seems that the reconstructed ones slightly
underestimate the correlations in the signal.  There are two
mechanisms leading to deviations of the construction from the
original: 
firstly, the result depends sensitively on the correct
estimation of the time scales involved in the original signal and 
secondly, the approximation of the pdf for the residence times by an
exponential is nothing exact, 
Nevertheless, the coincidence is remarkably good, taking into account
the possible error sources.  For real-world applications, one should
naturally calculate the pdf numerically and use the stored values.

\section{Discussion and conclusion}
\label{section:discussion}
In this paper we present a method to analyze and reconstruct
multi-time-scale signals, which originate either from extended (after
partitioning) or discrete systems. 
In our example, we use a simple partition of an extended system.  To
extract different time scales of the process, we apply the killing
window filtering technique to the signal. With the filtered signals we
use exit-time entropies to determine the Markovian order for the
stochastic description of the process under consideration. With the
information from the previous steps, we construct the multi time scale
signal. The reconstructed signal shows good statistical coincidence
with the original data.

We did not investigate in detail the question about the amount of data
needed for our method. This might vary from case to case, according to
the conditions that are needed for the respective application.  But
error estimates are standard in this case and we do not expect
novelties or surprises.  The procedure is consistent as a whole, there
are no ad-hoc assumptions necessary and everything is based purely on
observations.

More general scenarios with more than two time scales involved are
straightforward generalizations of the presented scheme, dependencies
at a deeper level can be included easily, e.g., dependence of a fast
time scale on the two preceding slower ones. This concept resembles
the slaving idea \cite{Ott-book-93,Haken-book-83} which proved to be a
powerful tool in pattern formation.

Applications are, to our opinion, mainly in the modeling and forecasting
of signals. One can well imagine such a stochastic model as a minimal
building block within a GCM or other simulation tools, which cannot
resolve fine-scale temporal motion due to numerical restrictions. In
this way, short-time fluctuations can be modeled by data analysis with
signals from experimental data.

As a stand-alone tool, one can use the obtained information to predict
the most probable track a tracer takes within a certain time, in this
case the reconstruction method is only good for supervising the
quality of the reconstruction.

It is hoped that these tools can be applied to to real data. We think
especially at data of the trajectories of Lagrangian tracers from
oceanographic experiments. There, we have to occupy with finding the
right partitioning of space. One can try to use ideas of dynamical
systems theory, as e.g. described in
\cite{Coulliette-Wiggins-2001}. These concepts however stand and fall
with the amount and quality of data.  If more information about the
system is available to yield additional reasoning for specific
partitions this should be used as input for a meaningful
coarse-graining.

\section{Acknowledgements}
We thank A. Vulpiani for exchange of ideas and valuable discussions.
Also M. Cencini, R. Pasmanter and D. Vergni have helped with interesting
remarks.  
M.A. and K.H.A. have been partially supported by the EU
network ``intermittency in turbulence'' (contract number
FMRX-CT98-0175). G.L. has received support and hospitality from the
University of Rome ``La Sapienza'', Department of Physics.
M.A. is currently supported by the DFG (German Research Foundation).

%%%%%%%%%%%%%%%%%%%%%%%%%%%%%%%%%%%%%%%%%%%%%%%%%%%%%%%%%%%%%%%

%\bibliographystyle{unsrt}
%\bibliography{markov.bib}

\begin{thebibliography}{10}

\bibitem{Cross-Hohenberg-93}
M.C. Cross and P.C. Hohenberg.
\newblock Pattern formation outside equilibrium.
\newblock {\em Rev. Mod. Phys.}, 65:851--1112, 1993.

\bibitem{Kantz-Schreiber-97}
H.~Kantz and T.~Schreiber.
\newblock {\em Nonlinear time series analysis}.
\newblock Cambridge University Press, Cambridge, 1997.

\bibitem{Gardiner-96}
C.W. Gardiner.
\newblock {\em Handbook of Stochastic Methods}.
\newblock Springer, New York, 1996.

\bibitem{Storch-Zwiers-99}
H.~v.~Storch and F.W. Zwiers.
\newblock {\em Statistical Analysis in Climate Research}.
\newblock Cambridge University Press, Cambridge, UK, 1999.

\bibitem{Cencini-Lacorata-Vulpiani-Zambianchi-99}
M.~Cencini, G.~Lacorata, A.~Vulpiani, and E.~Zambianchi.
\newblock Mixin in a meandering jet: a markovian approximation.
\newblock {\em J. Phys. Oceanogr.}, 29:2578--2594, 1999.

\bibitem{Murray-93}
J.D. Murray.
\newblock {\em Mathematical Biology}.
\newblock Springer, Berlin, 2nd edition, 1993.

\bibitem{Weisheimer-Dethloff-2001}
A.~Weisheimer, D.~Handorf, and K.~Dethloff.
\newblock On the structure and variability of athmospheric circulation regimes
  in coupled climate models.
\newblock {\em Atmospheric Sci. Lett.}, doi:10.1006/asle.2001.0034, 2001.

\bibitem{Corti-Molteni-Palmer-99}
S.~Corti and F.~Molteni T.~N. Palmer.
\newblock Signature of recent climate change in frequencies of natural
  atmospheric circulation regimes.
\newblock {\em Nature}, 398:799--802, 1999.

\bibitem{Coulliette-Wiggins-2001}
C.~Coulliette and S.~Wiggins.
\newblock Intergyre transport in a wind-driven, quasigeostrophic double gyre:
  An application of lobe dynamics.
\newblock {\em Nonlinear Processes in Geophysics}, 8(1-2):69--94, 2001.

\bibitem{Bower-91}
A.~S. Bower.
\newblock A simple kinematic mechanism for mixing fluid parcels across a
  meandering jet.
\newblock {\em J. Phys. Oceanogr.}, 21:173--180, 1991.

\bibitem{Samelson-92}
R.M. Samelson.
\newblock Fluid exchange across a meandering jet.
\newblock {\em J. Phys. Oceanogr.}, 22:431--440, 1992.

\bibitem{Chirikov-79}
B.V. Chirikov.
\newblock A universal instability of many-dimensional oscillator systems.
\newblock {\em Phys. Rep.}, 52:263--379, 1979.

\bibitem{ABCFVV-2000a}
M.~Abel, L.~Biferale, M.~Cencini, M.~Falcioni, D.~Vergni, and A.~Vulpiani.
\newblock Exit-time approach to $\epsilon$-entropy.
\newblock {\em Phys. Rev. Lett.}, 84(26):6002--6005, 2000.

\bibitem{ABCFVV-2000b}
M.~Abel, L.~Biferale, M.~Cencini, M.~Falcioni, D.~Vergni, and A.~Vulpian.
\newblock Exit-times and $\epsilon$-entropy for dynamical systems, stochastic
  processes and turbulence.
\newblock {\em Physica D}, 147:12--35, 2000.

\bibitem{note}
If one turns the procedure round, to come from infinite filter length, we first
  find a range of constant information production while we are inside the slow
  scale only and end with a steep increase in information production when we
  arrive at the fast scale.

\bibitem{Ott-book-93}
E.~Ott.
\newblock {\em Chaos in dynamical systems}.
\newblock Cambridge University Press, Cambridge, UK, 1993.

\bibitem{Haken-book-83}
H.~Haken.
\newblock {\em Synergetics}.
\newblock Springer, New York, 3rd edition, 1983.

\end{thebibliography}

%%%%%%%%%%%%%%%%%%%%%%%%%%%%%%%%%%%%%%%%%%%%%%%%%%%%%%%%%%%%%
\section{Appendix A}
\label{app:a}
The calculation of the block entropy in Eq. \ref{eq:h} is valid for
countable symbols. In this case, however the symbols are
not countable, as each symbol involves a continuous variable; the
residence time. The block entropy should therefore be calculated as
\begin{equation}
H^n = \sum \int d\tau \,p(S^n)\, \log(p(S^n)\;,
\end{equation}
where the sum extends over all possible permutations of n'th order.
For practical purposes, the residence times have to be binned, the
usual choice being a binning with resolution $\tau_r$. Then the block
entropy turns, after renumbering all the possible words, into the sum
$H^n(\tau_r)=\sum p(S^n)\, \log p(S^n)$, with summation over all
possible words now in in the now doubly discrete space. The entropy is
found as
\begin{equation}
h(\tau_r)^n = \lim_{n \to \infty} H(\tau_r)^{n+1} - H(\tau_r)^n\; .
\end{equation}
The limit of infinite time-resolution gives us the entropy {\it per
exit}, i.e.:
\begin{equation}
h^n = \lim_{\tau_r \to 0}h^n (\tau_r)\,.
\label{homegalim}
\end{equation}

In Ref.~\cite{ABCFVV-2000a,ABCFVV-2000b} more details are given,
including the topics of continuous signals and several rigorous bounds
for the entropy together with some applications.

%%%%%%%%%%%%%%%%%%%%%%%%%%%%%%%%%%%%%%%%%%%%%%%%%%%%%%%%%%%%%%%%%%%

\end{document}